# Integrated Nanophotonics Architecture for Residue Number System Arithmetic

Jiaxin Peng[1], Shuai Sun[1], Vikram K. Narayana[1], Volker J. Sorger[1], Tarek El-Ghazawi[1,*]

[1]Department of Electrical and Computer Engineering, George Washington University, 800 22nd St NW, Washington, DC 20052

[*]Corresponding author: tarek@gwu.edu

**Abstract**

Residue number system (RNS) enables dimensionality reduction of an arithmetic problem by representing a large number as a set of smaller integers, where the number is decomposed by prime number factorization using the moduli as basic functions. These reduced problem sets can then be processed independently and in parallel, thus improving computational efficiency and speed. Here we show an optical RNS hardware representation based on integrated nanophotonics. The digit-wise shifting in RNS arithmetic is expressed as spatial routing of an optical signal in 2×2 hybrid photonic-plasmonic switches. Here the residue is represented by spatially shifting the input waveguides relative to the routers outputs, where the moduli are represented by the number of waveguides. By cascading the photonic 2×2 switches, we design a photonic RNS adder and a multiplier forming an all-to-all sparse directional network. The advantage of this photonic arithmetic processor is the short (10's ps) computational execution time given by the optical propagation delay through the integrated nanophotonic router. Furthermore, we show how photonic processing in-the-network leverages the natural parallelism of optics such as wavelength-division-multiplexing or optical angular momentum in this RNS processor. A key application for photonic RNS is the functional analysis convolution with widespread usage in numerical linear algebra, computer vision, language- image- and signal processing, and neural networks.

## I. Introduction

The residue number system (RNS) in the field of digital computer arithmetic has been proven to have advantages in decomposing the larger integers into a set of smaller integers in calculation and performs the calculations independently and in parallel.

Fundamentally, adapting photonics into the RNS for signal process could benefit from: (1) the fast execution time which is given by the photon's time-of-flight through the structure, (2) the nature of light that a photon always has to propagate with a momentum which in other words, the algorithms could achieve date processing complies with the data propagation and (3) In addition, the wave division multiplexing (WDM) capable with broadband nanophotonic devices could even adapt parallelism into RNS for higher bandwidth and energy efficiency.

In RNS, an integer number is represented as a set of residues obtained using a set of moduli. RNS arithmetic is executed digit-wise without carry propagation, a large number can be decomposed into smaller ones that can be operated on in parallel [1]. In RNS, a large integer is represented using a set of smaller integers so that computation may be performed more efficiently. RNS is a fast and highly-scalable technique for high-performance computing [2]. Each RNS arithmetic operation can be decomposed into a number of sub-operations that can be carried out by multiple circuits in parallel. With a suitable number encoding scheme, these sub-operations can be achieved by bit shifts, thereby making RNS computation amenable to optical realization. In addition, the data movement of RNS arithmetic, allowing the use of optical switch with high performance. Optical communication scales more favorably than electronics in signaling distance in terms of data bandwidth and delay, due to the lack of capacitive parasitics [3,4]. However, the promise of optical computing has not delivered compute engines that outperform electronics when considering bandwidth, energy-per-bit, system size, and cost [5]. Fundamental physics, however, points towards a higher information density of photonics since multiple properties such as amplitude, phase, orbital angular momentum, polarization, long-range entanglement, can be utilized simultaneously [6,7]. Here we proposed a hybrid photonic-plasmonic (HPP) RNS adder with all-to-all sparse directional (ASD) schematic, based on cascaded

HPP 2×2 switches forming a crossbar with broad spectrum operating bandwidth. Optical computing inherently brings with it the extra benefits of higher rates of optical transmission. Moreover, optical realization yields very low propagation losses. Thus, a chain of RNS operations to be cascaded together with a negligible increase in compute time, a feat not possible using conventional electronics. This method opens the door of a new class of computing in switching. The proposed RNS could be developed as RNS multiplier with little change. Moreover, convolution computation units could be implemented easily by these two basic RNS units.

## II. Residue Number System

The RNS uses remainders of different moduli to describe a given number. For a single modulo $M$, the number $X$ has a specific residue, or a remainder, $r$. If X = 96, M = 11, then $r = |96|_{11} = 8$. Assuming there are a set of moduli $\{M_1, M_2, …, M_n\}$ then the corresponding residues of $X$ are $\{r_1, r_2, …, r_n\}$. The moduli $\{M_1, M_2, …, M_n\}$ should not have a common factor with each other. Typically, prime numbers are chosen as moduli to avoid any possible common factors among the moduli set. RNS could represent numbers ranging from 0 to $(\prod_{i=1}^{M} M_i - 1)$. In RNS, $X$ could be represented as $\{r_1, r_2, …, r_n\}_{[M1, M2, …, Mn]}$. For example, in a residue number system with moduli $\{11, 19, 23\}$, X = 96 and Y = 32 could be represented as $\{8, 1, 4\}_{[11, 19, 23]}$ and $\{10, 13, 9\}_{[11, 19, 23]}$ respectively.

Within the representation range, each number is uniquely represented by the set of its residues. Since computations are carried out for each modulus independently, RNS calculation does not require any carry propagation. It thus turns out that operations on larger numbers can be decomposed into smaller number calculations, resulting RNS as a fast and easy to parallelize. Also, it provides the possibility of computing in network.

*Example 1:* X + Y = $\{8+10, 1+13, 4+9\}_{[11,19,23]} = |18|_{11}, |14|_{19}, |13|_{23} = \{7, 14, 13\}_{[11,19,23]}$

Above example shows the addition of operation *96+32* in RNS. It applies to subtraction and multiplication as well. For the residue number system, the arithmetic operations of addition and multiplication may be executed in the same time as required for an addition operation. The main difficulty of the residue number system relative to arithmetic operations is the determination of the relative magnitude of the two numbers expressed. The residue code is of little utility for general purpose computation, but the code has many characteristics which recommend its use for special purpose computations. In this paper, we focus on single residue arithmetic. But it is very easy to expand to RNS since it is digit-wise.

## III. RNS Adder Schematic

The conceptual idea of RNS computing modules based on different types of 2×2 optical switches have been proposed, however, only based on a mesh grid of switches [8]. This design is potentially integrated into optical circuit with low power consumption. However, the area requirement will be high based on their schematic. Therefore, we proposed a new design based on a strict-sense non-blocking router design with our HPP ITO switch from previous work [9,10]. It required fewer optical switches for each modulo system. The tradeoff is complicated electrical control circuit requirement since each switch will have one separate control signal now. Also, a look-up table (LUT) is necessary to store the states of all the optical switches. However, it turns out our design requires less area than the mesh RNS adder design. Our HPP switches gain benefits of energy consumption and propagation time compared to MRR, MZI and all-optical-switch. Also, we point out the possible future application on our design to implement convolution operations.

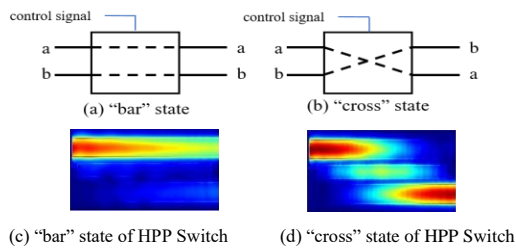

*Figure 1 Two States of a 2×2 Switch. (a) and (b): the conceptual schematic of two states in a 2×2 switch. (c) and (d): the top view of the FDTD simulation results of two states in our 2×2 Hybrid Photonic-Plasmonic switch with two silicon waveguides (up and down) as buses and a switching island covered by indium tin oxide in between to achieve signal switching. [9]*

A.Tai et al proposed a conceptual design of optical RNS computing, including adder and multiplier [8] based on one-hot-encoding. Each location represents a specific number in a modulo-$M$ system, and thus needs $M$ bits. Both RNS addition and multiplication could be implemented by using multiple 2×2 optical switches. Fig.1 shows an essential 2×2 optical switch. Based on a control signal, the two output signals could be either one of the input optical signal. Hence, there are two states. Fig.1(a) allows the light to go through on the same side is called *bar* state which has the bias voltage on the switching island in the middle, while Fig.1(b) is called *cross* state without any biasing voltage which flips the light to the other side of the bus. In bar state, two bunches of light will go straight respectively; in cross state, the input lights will go cross each other.

In the design by Tai. et al, the optical I/O signal is based on one-hot-encoding, which means once the specific location has a signal, there is the corresponding input/output. A modulo-$M$ number thus requires $M$ bits for representation, and only a single bit is "1" at any time. The use of one-hot encoding allows the arithmetic to be carried out by just bit shifts, since the input and output are both represented by a single active signal in the $M$ bits. For RNS addition, the two summands are separated to optical input and a control signal respectively. By changing the control signal row by row, the optical signal could be routed to any desired output. The control signal of each row is stored by SR-Flip-flops. Finally, there is several photo-detectors are placed at the end of each waveguide for identifying results. Bakhtiar et al designed a modulo-5 adder and system by an ultimate fast optical switch [11] based on Tai et al's conceptual schematic. The design of Leily A. Bakhtiar et al required $M\times(M-1)$ switches for adder as well. Though the above idea is designed on optical switch, Bakhtiar et al indicated that transistors could also implement the RNS adder. While applying the control signals to drive the gate, the other summand is entered from the drain of transistors, and finally results are coming from the source of transistors. It seems electrical version is good since transistors are small and easy to implement; however, as modulo-$M$ system goes up, the distance between control signal and SR-flip-flop, where storing the control signal, will increase exponentially. The reason is that each wire should be charged. Furthermore, RC delay and power consumption increase as well. These drawbacks limit the transistor implementation in this design schematic, especially in large modulo-$M$ design.

The other drawback in the RNS adder design of Leily A. Bakhtiar et al, this method is unfeasible since all optical switches uses higher order nonlinearities which usually requires $MW/cm^2$ level of incident energy. This tremendous energy overhead makes the all optical switch hard to be integrated on-chip or up-scaling for large RNS. Moreover, depending on the fact that RNS addition and multiplication are one-to-one relationship, which acts like all-to-all routing, $M\times(M-1)$ switches for a m-to-m router is extremely high. Therefore, a novel schematic of RNS adder is proposed, called all-to-all sparse directional (ASD) schematic. Based on a set of parallel waveguides, several 2×2 switches are built on them. By controlling the control signal of each switch, the light could go as desired path. The photodetector translates the result of an operation based on the position. This RNS adder is programmable since the control signal could be reprogramed as demand.

Previously, S. Sun et al proved that a n×n router needs $(M-1)^2/2$ 2×2 optical switches to implement a wide sense communication (if n is odd) [9]. In our case, $M$ is prime so that $M$ is odd except for 2. To ensure strict non-blocking communication, two more optical switches are needed. Therefore, our adder requires $(M-1)^2/2+2$ optical switches for M>2.

Fig.2 shows the proposed optical RNS adder design of modulo-5 system. Obviously, the setting time is required. SR-Flip-Flops are used to synchronize control signal, which means one optical switch is connected to a single SR-Flip-Flop. The whole system is one-hot encoding. Each light beam at specific location represents one corresponding number. By setting up the states of each switch, no matter what another summands or factor is, the input light beam will go over the desired path, and the result could be detected by the photodetector. Red line in Fig.2 represents the operation of *2+4*. By switching the states, the 2×2 switch could lead the input light to corresponding position by searching the LUT. As for Leily A. Bakhtiar et al's design, a modulo-5 adder requires $5 \times 4 = 20$ switches. We saved 10 switches on a single modulo-5 adder. For the whole RNS system, a large number of single modulo system will be needed. It will be a huge advantage in terms of area and energy.

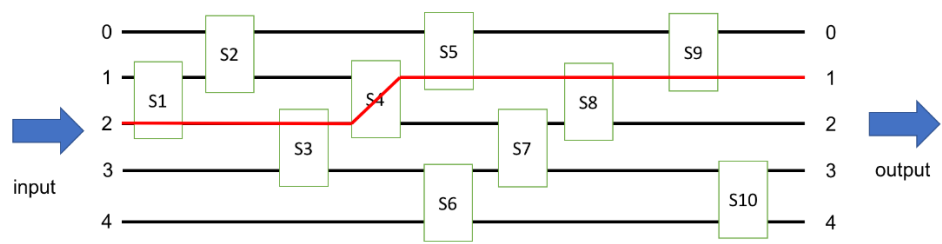

*Figure 2 Modulo-5 RNS Adder with example 2+4 = |6|$_5$ = 1*

## IV. Hybrid Photonic-Plasmonic Switch

The actual challenge of the aforementioned photonic devices is the fundamentally weak light matter-interaction (LMI) originating from the small dipole moment of the optical wave acting at the matter atom, which leads to 10-100's of micrometer long interaction lengths for optoelectronic devices [12]. However, making the photon more polaritonic (matter-like), such as in plasmonics, enables strong LMIs and hence short devices, which has positive effects on the device performance of the device [5,13]. Positive effects of wavelength-scale active opto-electronics are a) low electrical capacitance, b) short photon lifetimes allowing rapid re-excitation of the device, and c) high energy efficiency due to the small capacitance and driving voltage. Nevertheless, the high intrinsic ohmic loss limits the plasmonic devices for large-scale integration [3,14].

Based on the reasons mentioned above, combining the low propagation loss silicon photonic waveguides with ultra-fast tunable material enabled plasmonic components becomes a possible solution to alleviate the fundamental drawbacks from both sides. This hybrid photonic-plasmonic (HPP) device is able to combine high LMI active optoelectronics with low-loss passive photonic elements as we demonstrated in our previous work [9,10]. With a tunable indium tin oxide layer deployed in the middle switching island, the light input is able to be either routed to the other side of the bus (cross state) or kept in the same bus (bar state) based on the bias voltage as a $2 \times 2$ switch (Fig.1(c), 1(d)). As the building block of our RNS model, this HPP switch provides up to 200 GHz switching time and femtojoule level switching energy which are critical for high efficient data processing. Moreover, the 13 $\mu m^2$ footprint also gives the potential for dense integration networks-on-chip [15-17].

## V. Performance Comparison

To implement the conceptual $2 \times 2$ switch, micro-ring (MRR), Mach Zehnder interferometer (MZI), all-optical-switch (AOS), in addition to our proposed HPP switch, could be the potential method. Based on the two different schematics of RNS adder with four types of $2 \times 2$ switches, Table 1 shows the comparison of number of optical components and control units, energy consumption, area, as well as speed based on single operation.

Although our proposed schematic requires complicated electrical control circuits with a LUT, the actual area needed is relative small due to the actual size of transistor is quite smaller than a single optical switch. As shown in Table 1, for a modulo-5 adder, the area and power consumption of control circuits is around 2.5 times of the mesh grid RNS adder design. All-optical-switch is not being use anymore by its huge switching energy consumption, despite of its compact size and terahertz level switching speed. MRR consumes less energy but requires very large area and response time, while MZI needs less footprint and propagates faster with higher energy consumption. Our HPP switch works better. HPP ITO switch with our proposed schematic is just double size as MZI, but requires only 10% energy consumption that MZI needs to reach the same propagation time per device, which is 0.1ps only. Our HPP ITO switch works better than MRR does in terms of area, speed, and energy consumption.

| Parameters | | Mesh RNS Model [8] | | | | ADS RNS Model (proposed) | |
|---|---|---|---|---|---|---|---|
| | | MMR [18] | MZI [19] | AOS [19] | Scale with $M$ | HPP Switch | Scale with $M$ |
| # of Component | | 20 | 20 | 20 | M(M-1) | 10 | $(M-1)^2/2+2$ |
| # of Control Circuit | | 4 | 4 | 4 | M-1 | 10 | $(M-1)^2/2+2$ |
| # of Look-up Table (LUT) | | - | - | - | - | 50 | $M[(M-1)^2/2+2]$ |
| Energy/op. | Thermal | 13.8 fJ/bit | - | - | M(M-1) | - | - |
| | Switching | | 500 fJ/bit | 12 nJ/bit | M(M-1) | 5.2 fJ/bit | $(M-1)^2/2+2$ |
| | Control | 0.8 fJ/bit | 0.8 fJ/bit | 0.8 fJ/bit | M-1 | 2 fJ/bit | $(M-1)^2/2+2$ |
| Area | Component | 3200 $\mu m^2$ | 200 $mm^2$ | 10 $\mu m^2$ | M(M-1) $\times A_{device}$ | 200 $\mu m^2$ | $[(M-1)^2/2+2] \times (A_{device}+ A_{control})$ |
| | Control Circuit | | 0.8 $um^2$ | | + (M-1) $\times A_{control}$ | 2 $\mu m^2$ | + $M[(M-1)^2/2+2] \times A_{LUT}$ |
| Response Time | | 40 ps | 14.3 ps | 0.2 ps | 1 | 5.1 ps | 1 |
| Per device prop. time | | 0.754 ps | 0.1 ps | 0.01 ps | M-1 | 0.1 ps | M |

*Table 1 Comparison of Mesh Grid RNS Model and ASD RNS Model (proposed) in Modulo-5 Adder with microring resonator (MMR), Mach Zehnder interferometer (MZI), All-Optical switch (AOS), and hybrid photonic-plasmonic (HPP) ITO switch and their scale with M. The LUT is the look-up table that needed by the ADS RNS model.*

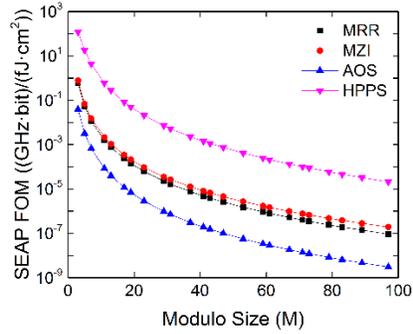

*Figure 3 Speed (Energy Efficiency) (Area Efficiency) Product comparison of the mesh RNS model and the router-based RNS model on 1) micro-ring resonator, 2) Mach Zehnder interferometer, 3) all optical switch, and 4) hybrid photonic-plasmonic switch as building blocks based on the parameters from Table 1. (MRR = micro-ring resonator, MZI = Mach-Zehnder interferometer, AOS = all-optical switch, HPPS = hybrid photonic-plasmonic switch).*

Fig. 3 shows the Speed-Energy-Area Product FOM versus the increasing modulo size with four types of switches. Combing HPP ITO switch with proposed schematic, SEAP FOM performs 100 times better than MRR and MZI do. Our design requires less area; our switch requires less energy with higher speed. As the modulo size increase, overall performance decreases. Since we have more components, the size and power requirement should increase as well. We keep our advantage when compared to other switches with the mesh schematic RNS adder.

## VI. Application

Based on this tiny RNS adder, we developed a similar structure for RNS multiplier as well, shown as Fig.4. The multiplier is slightly different from the adder. Once one of the factor is zero, the result should be zero. Thus, a several switches on the input *0* could be saved. However, considered the situation that other summand is *0*, before each non-zero output is compatible with additional output zero. To avoid the cross over, several waveguides are added. It is also programmable since the states of each optical switches could be changed accordingly. By combining both adder and multiplier, several applications are enabled such as convolution for signal processing, or neural networks.

To the best of our knowledge, electrical and the existing optical RNS adder design calculate only one operation at a time. According to [11] and [21], the efficiencies of both designs are *1/M* for a single operation. Though RNS optical computing runs fast, it is still inefficient since most of the resources are idle. Another advantage of the HPP device is it is wavelength division multiplexing (WDM) capable, which allows multiple bunches of light with different wavelength propagate in the device at one time. Therefore, we propose a new scheme to allow a few RNS calculations are being operated concurrently as Fig.5 shown. By adding ring resonators and photo-detectors at the end of output, light could be identified to corresponding operations. Once the digital control signal is set, input a broad-band light with multiple wavelength $\lambda_1$ $\lambda_2$ $\lambda_3$... $\lambda_v$ could be represented to different operations. For example, once we decide one of the summand is 4, and it is an adder, then now our model allow two situations: 1) same input summand with different wavelength. Two "1"s are input now. The photo-detector at the output is able to identify corresponding results for two operations with frequency of $\lambda_1$ and $\lambda_2$ (green and blue lines in Fig.5 respectively). The MRR with photodetector will recognize the result of both operation 1 and 2 are 0. 2) different input summand. Operation *3* "0+4" has frequency of $\lambda_3$ (purple line in Fig.5). Purple line shows finally the photo-detector of $\lambda_3$ will recognize the result of operation *3* is 4. Above shows that our design is WDM capable, allowing multiple calculations operates simultaneously. Then *M* ring resonator with photo-detector could detects specific wavelength to know the results of corresponding operation. Moreover, if the control signal is constant, a few operations could be calculated at the same time.

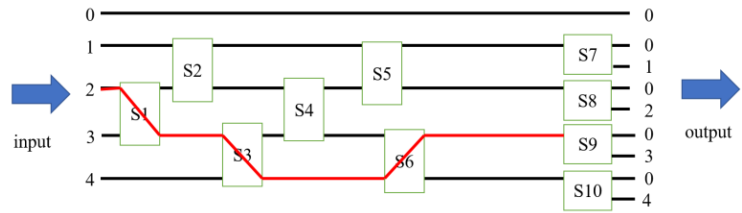

*Figure 4 Modulo-5 RNS Adder with example $2 \times 4 = |8|_5 = 3$*

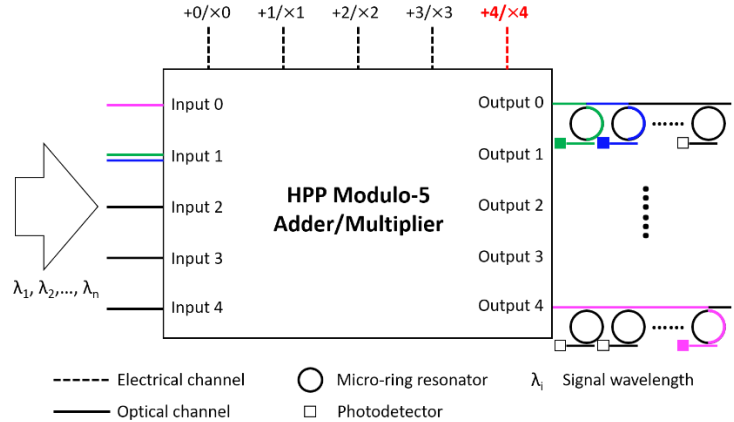

Figure 5 Schematic of WDM Modulo-5 RNS Computation Unit.

Overall, our design is suitable for specific type of operations. For instance, in convolutional neural network, 90% of the calculation is the multiplication-accumulation-calculation (MAC) operation [22]. A same weight matrix will be calculated million times. After setting the control signal same as the weight, the MAC operation could be calculated very fast, resulting from both the WDM design, as well as the possibility of high frequency of optical computing.

## VII. Conclusion

Residue number system satisfies the future trend for large number calculation since it could be decomposed into small numbers. More importantly, the RNS operations are digit-wise, allowing computing in parallel. Optical computing for RNS provides possible higher frequency of optical transmission. By proposing the new schematic with parallel waveguides, the design requires less area. Although complex control circuits and algorithm are required, the smaller size is extremely critical in real world. Also, we proposed a HPP ITO switch which is suitable for RNS design. Comparing with MRR, MZI and all-optical switch, our proposed design works 100 times better in terms of SEAP FOM overall. It requires less area and energy for faster speed.